\documentclass[10pt]{article}
\usepackage{epsfig}
\usepackage{apjfonts}
\usepackage{emulateapj}

\righthead{Imaging of HD 141569 Disk}
\lefthead{Weinberger et al.}

\def\micron{$\mu$m}

\def\degr{$^\circ$}

\begin{document}

\submitted{Accepted 3 Sep 1999 for publication in ApJ Letters}
\title{The Circumstellar Disk of HD 141569 Imaged with NICMOS}

\author{A. J. Weinberger\altaffilmark{1}, E. E. Becklin\altaffilmark{1},
G. Schneider\altaffilmark{2}, B. A. Smith\altaffilmark{3},
P. J. Lowrance\altaffilmark{1}, M. D. Silverstone\altaffilmark{1},
B. Zuckerman\altaffilmark{1}, and R. J. Terrile\altaffilmark{4}}

\altaffiltext{1}{Department of Physics and Astronomy, University of
California Los Angeles, Box 156205, Los Angeles, CA 90095; \\
alycia,becklin,lowrance,murray,ben@astro.ucla.edu}

\altaffiltext{2}{Steward Observatory, University of Arizona, 933
N. Cherry Ave., Tucson, AZ 85721; gschneider@as.arizona.edu}

\altaffiltext{3}{Institute for Astronomy, University of Hawaii,
Honolulu, HI 96720; brad@mahina.ifa.hawaii.edu}

\altaffiltext{4}{Jet Propulsion Laboratory, MS 183-503, Pasadena, CA
91109; Richard.J.Terrile@jpl.nasa.gov}

\begin{abstract}
Coronagraphic imaging with the Near Infrared Camera and Multi Object
Spectrometer on the Hubble Space Telescope reveals a large, $\sim$400 AU
(4$''$) radius, circumstellar disk around the Herbig Ae/Be star HD
141569.  A reflected light image at 1.1 \micron\ shows the disk oriented
at a position angle of 356 $\pm$ 5\degr\ and inclined to our line of
sight by 51 $\pm$ 3\degr; the intrinsic scattering function of the
dust in the disk makes the side inclined toward us, the eastern side,
brighter.  The disk flux density peaks 185 AU (1.$''$85) from the star
and falls off to both larger and smaller radii.  A region of depleted
material, or a gap, in the disk is centered 250 AU from the star.  The
dynamical effect of one or more planets may be necessary to explain this
morphology.
\end{abstract}

\keywords{circumstellar matter --- stars: emission line --- stars: individual
(HD 141569)}

\section{Introduction}

The discovery of disks around main-sequence stars has opened
possibilities for studying nascent planetary systems.  Old Herbig Ae/Be
and $\beta$~Pic analog stars lie at a transitional age between
pre- and zero-age main sequence stars.  Both classes have infrared
excesses indicative of the presence of warm circumstellar dust as well
as luminosities and colors consistent with those of young stars.  Their
circumstellar material, in the form of disks, may be composed of both
remnant proto-stellar and secondary ``debris'' dust generated by
collisions.

HD 141569 (= IRAS 15473-0346 = SAO 140789), classified as B9.5Ve by
Jaschek \& Jaschek (1992), is a Herbig Ae/Be star with double peaked H$\alpha$
emission.  Jaschek, Jaschek, \& Egret (1986) first identified it as a main sequence star
with infrared excess and an SED peaking at $\sim$60 \micron.  The disk
has an optical depth $\tau_{3-100\mu m} = \rm{L}_{disk}/\rm{L}_{*} = 8.4
\times 10^{-3}$ (\cite{Zuckerman95}), which is a few times larger than
the two well-known infrared excess stars, $\beta$~Pic and HR 4796A.
With an Hipparcos measured distance of 99 $^{+9}_{-8}$ pc (so 1$''
\approx$ 100 AU), the star may lie on the outskirts of the dark cloud
complex L134/L134N (\cite{Sahu98}).  Assuming its B$-$V = 0.095 mag color
is indicative of interstellar reddening, HD 141569 has an absolute
magnitude M$_V$=1.6 mag (van den Ancker et al. 1998).  Such a low
luminosity for its spectral type is consistent with that of other young
main sequence A stars (\cite{Jura98}).

The age of HD 141569 is not well determined.  It is located
approximately on the zero age main sequence, has an infrared optical
depth similar to that of other young A stars, has strong CO emission
(\cite{Zuckerman95}), and has two T-Tauri like companion stars
(\cite{weinberger99b}).  Together, these constrain its age to
10$^6-$10$^7$ yr.

The 1.1\micron\ results reported here are part of our guaranteed time project
to study the environments of nearby stars with the Near Infrared Camera and
Multi Object Spectrometer (NICMOS) aboard the Hubble Space Telescope (HST)
(\cite{schneider98a}).  HD 141569 was included because of its large
far-infrared optical depth, CO emission, and extended elliptical disk in
mid-infrared images by Silverstone et al. (1999).  The scattered light disk around
HD 141569 has also been observed with NICMOS at 1.6 \micron\ by
Augereau et al. (1999a).

\section{Observations}

Coronagraphic observations of HD 141569 were made on 27 September 1998
with NICMOS Camera 2. The F110W filter ($\lambda_{central}$=1.10
\micron, FWHM=0.59 \micron) was used to maximize the angular
resolution while minimizing scattered light.  With the target positioned
in the 0.$''$3 radius coronagraph, three MULTIACCUM sequences were
obtained, for a total integration time of 607.9 sec at each of two
spacecraft orientations, rolled 8\degr\ about the target axis with
respect to each other.  To minimize time-dependent point spread
function (PSF) variations, the two image sets were obtained within 40
minutes.

Prior to each set of coronagraphic observations, 0.36~s target
acquisition frames were obtained with the F171M
($\lambda_{central}$=1.72 \micron, FWHM=0.07 \micron) filter.
Contemporaneous lamp flats and backgrounds were obtained at F160W for
the purpose of locating the coronagraphic hole and enabling good
flat-fielding near the hole.

\section{Data Analysis}

The MULTIACCUM data sets were processed with the Nicreduce software
(\cite{MacLeod97}).  Dark frames taken on 20 September 1998 were used to
subtract the dark current and correct the detector shading.  A special
coronagraphic flat field was created by modifying a standard high S/N
on-orbit F110W reference flat field. Within a radius of 12 pixels
(0.$''$91) of the hole, the pixel sensitivities were measured in the
F160W hole-finding image, scaled to F110W, and replaced in the reference
flat.  After bias subtraction, linearization and flat fielding, known
bad pixels were corrected with a distance weighted average within a
radius of eight pixels, and the three MULTIACCUM images from each
spacecraft orientation were medianed.  The best available photometric
calibration was applied to the final F110W images in which 1 ADU
s$^{-1}$ = 2.031$\times$10$^{-6}$ Jy and 0~mag = 1775 Jy
(\cite{riekecal}).

In the reduced and calibrated NICMOS images, scattered light from the
star still dominates the flux levels at radii of 1---5$''$, so in order
to see the disk, a PSF star must be subtracted.  The same observing
strategy and the F110W filter were used for 14 other stars in this GTO
program, and targets which do not show extended disk emission form a
library of possible PSF stars.  The detailed character of the NICMOS PSF
changes with time primarily due to thermal variations in the HST optical
assembly (\cite{kulkarni99}); hence some PSF star subtractions produce
much lower residuals than others.  Each PSF image was registered using
cubic convolution interpolation to both orientations of HD 141569.  Only
stars brighter than HD 141569 were considered, and the PSF subtraction
residuals are dominated by systematic uncertainties.

Scattered light from two nearby stars outside the field of view, see
\S\ref{diskmorph}, pollutes the northwest corner of the image.  For both
roll angles, we used synthetic PSFs generated with Tiny Tim
(\cite{Krist}) and positioned them at the locations of the two stars to
subtract this excess scatter.  A small amount of residual flux, however,
can still be seen in Figure \ref{figimg}.

Figure \ref{fig_radialavg} compares the azimuthally averaged disk flux
observed after subtraction of each of the three best matched PSF stars.
Two of these PSF stars, $\tau^1$ Eri and 49 Cet, were disk candidates in
which no disk was detected, and the third, HR 4748, was observed
specifically as a PSF star.  At radii greater than 1.$''$6 (160 AU), the
morphology of the disk is consistent in all three subtractions.
Subtraction of 49 Cet yielded the smallest amplitude residuals, and all
further analysis was carried out on images using it as the PSF.

The PSF-subtracted images from each telescope roll were rotated to a
common orientation and averaged.  Regions of the image obscured by the
diffraction spikes in one roll were replaced where possible by uncorrupted
regions in the other roll.

\section{Results} \label{results}

The PSF subtracted, roll-combined coronagraphic image of HD 141569 is
shown in Figure \ref{figimg}a.  An elongated disk is clearly evident at
radii from $\sim$1.6 to 4$''$ (160 to 400 AU) from the star.  Elliptical
isophotes were fit to the disk image yielding a position angle of 356
$\pm$ 5\degr\ and ratio of the minor to major axes of 0.63 $\pm$ 0.04.
If, as is likely for dust orbiting a star, the disk is intrinsically
circular, the axial ratio implies an inclination of 51$\pm$3\degr\ from
face-on.  This geometry was used to ``deproject'' the disk to a face-on
view (see Figure \ref{figimg}b).

The azimuthally averaged disk surface brightness, shown in Figure
\ref{fig_radialavg} as computed from the deprojected image, peaks at 0.3
mJy arcsec$^{-2}$ at a radius of 185 AU and falls off to larger and
smaller radii.  Fitting the slope of the surface brightness with a power
law gives r$^{-3.2}$ for 190 $<$ r $<$ 250 AU.  At 250 AU an annulus of
lower surface brightness, or a ``gap'' in the disk, can be seen, which
is apparent in Figure \ref{fig_tauomega} where the surface brightness
has been multiplied by 4$\pi$r$^2$ to determine the surface
density of scatterers (see \S\ref{grains}).  On either side of this gap,
the surface density of scatterers does not vary appreciably.  At
r$\gtrsim$340 AU, the surface brightness of the disk falls off more
steeply, as r$^{-5.9}$.

The disk flux density, in a rectangle of area 63 sq.~arcsec outside of 0.$''$6
from the star and extending to 5.$''$3 from the star along the major axis of
the disk, is 7.2$\pm$1.8~mJy.  This integration excluded pixels in the
diffraction spikes; we make a purely geometric correction by finding the
average flux density at every radius and multiplying it by the number of
excluded pixels.  The corrected total flux density is 8 $\pm$ 2 mJy.  The
F110W magnitudes of 49 Cet and HD 141569 were extrapolated from their H-band
magnitudes, and the uncertainty in their flux ratio completely dominates the
uncertainty in the photometry of the disk.  A change of $\sim$2\% in the
subtracted PSF flux, makes a $\sim$20\% difference in the integrated disk flux
density.

The eastern side of the disk is 1.5$\pm$0.2 times brighter than the
western side.  This difference is not an artifact of the PSF subtraction
and can be explained by dust particles which preferentially scatter in
the forward direction.  Such phase functions apply even when the
particles are a few times larger than the wavelength of the scattered
light (\cite{vandehulst}) and are produced in recent models of fluffy
grains (e.g. \cite{Wolff98}).  Small particles probably do exist in the
HD 141569 disk (see \S\ref{grains}).  The brighter side, therefore, is
inclined toward us, and the ratio of forward to backward scattering
implies an asymmetry parameter, $g$, of 0.11 in the analytic phase
function of Henyey \& Greenstein (1941).  This is consistent with the results of
Augereau et al. (1999b) on HR 4796A in which the upper limit for $g$ was
0.15.

\section{Discussion}

\subsection{Disk Morphology} \label{diskmorph} \label{gapsection}

The surface brightness of the disk has a complicated radial morphology
which cannot be represented by a single power-law.  The density of
scatterers, assuming the disk is everywhere optically thin, is as high
at radii up to about 360 AU from the star as it is in the region 200 AU
from the star.

Dynamical processes could play a large part in shaping the disk.  HD
141569 may be the brightest member of a multiple star system, since two
bright stars (henceforth ``B'' and ``C'') with H = 8.6 and 9.4 mag lie
7.$''$57 ($\gtrsim$750 AU) and 8.$''$93 ($\gtrsim$900 AU) away at
respective position angles of 311.5\degr\ and 310.0\degr.  The high
probability that these stars are physical companions and assessments of
their ages are discussed in a separate paper (\cite{weinberger99b}).  If
the three stars are physically associated, the two companions would
likely have a significant dynamical effect on the disk.  If the
companion stars are in the plane of the disk, the physical separation of
``A''--``B'' is 990 AU and that of ``B''--``C'' is 190 AU.  Then, the
ratio of the semi-major axes of the wider to the closer pair would be
only $\sim$5.2.  This is not expected to be a stable triple system
(\cite{Eggleton95}), although such young stars may not yet have had time
to become unbound.

If the companions are out of the disk plane, they could be much further
from the primary, and their dynamical effect could be to excite
significant velocities in the disk dust perpendicular to the plane.
This would puff up the disk and possibly flare it. Increased collision
velocities could promote the breakup of larger particles into smaller
and could help account for the large number of scatterers far out in the
disk.

The dip in surface brightness at r$\approx$250 AU which is most easily
seen along the major axis (Figure \ref{figimg}), is a striking feature
of the disk.  The full width at half minimum of this gap is 60 AU, to
which the intrinsic size of the NICMOS PSF contributes $\sim$5\%.  The
depth of this dip in terms of the density of scatterers ($\tau\omega$;
Figure \ref{fig_tauomega}) is $\sim$14\% of the average value between
185 and 340 AU.

No point source is seen within the disk, and to constrain the
luminosities of possible companion bodies, we planted PSF ``stars,''
generated with Tiny Tim (\cite{Krist}), at random locations in the image
at one telescope roll angle subtracted from the image at the other roll.
The limiting magnitude (3$\sigma$) of an object which could be detected
at each radius is plotted in Figure \ref{fig_srclimit}.  This limit does
not apply to the region obscured by the diffraction spikes which, at the
radius of the gap, hide $\sim$17\% of the disk.

At the radius of the gap, the limiting magnitude of 20.3 and an assumed
age $\leq$10 Myr correspond to a planet with mass $\leq$0.003 M$_\odot$
(3 M$_{\rm JUP}$) according to cooling curves by Burrows et al. (1997).  At
this age, the brown dwarf-planet transition occurs at
log(L/L$_\odot$)$\approx -$3 which corresponds to an F110W magnitude of
$\sim$16. Figure \ref{fig_srclimit} shows that either a star or brown
dwarf would be clearly detectable in our image at any radii greater than
$\sim$80 AU.

If the gap is cleared only by the gravity of a co-orbital planet, the
planet mass is: $M/M_\odot = C (\Delta a/a)^3$ where $\Delta a$ and $a$
are the width and radius of the gap, respectively, and the constant
$C\approx$ 0.1 (\cite{Lissauer}).  This corresponds to a planetary mass of
$\sim$1.3 Jupiters, which given the detection limits discussed in
\S\ref{gapsection}, could not have been detected.  The orbital period
for material 250 AU from the 2.3 M$_\odot$ star is 2600 yr. If we assume
that it takes $\gtrsim$300 orbital periods to clear material in a gap,
based on models of disks by Bryden et al. (1999), then a gap can be opened
in $\gtrsim$8$\times$10$^5$~yr.  That the gap is not completely clear
may mean that processes are moving particles through the disk.  The
Poynting-Robertson timescale for particles $\gtrsim$1\micron\ is
$\sim$4$\times$10$^6$~yr, similar to the age of the star.  Radiation
pressure will preferentially clear small grains on a timescale of
$\sim$10$^4$~yr and could fill in a gap if such grains are present, or
being created, in the disk.

\subsection{Constituent grains}\label{grains}

We can combine the information we infer from the NICMOS images with what
is known about the disk from mid to far-infrared measurements.  For
reflection from optically thin dust, $\tau\omega = 4\pi\phi^2 \,
\frac{S}{F}$, where $\phi$ is the angular distance of the scatterers
from the illuminating star, $S$ is surface brightness, $F$ is the
received flux from the star, $\tau$ is the optical depth of scatterers,
and $\omega$ is the albedo (Figure \ref{fig_tauomega}).  At the peak
surface brightness, $\phi$=1.$''$9, $S$=0.3 mJy arcsec$^{-2}$, and $\tau
\omega = 4 \times 10^{-3}$.  The optical depth to visual absorption
implied by the far infrared excess is $8.4 \times 10^{-3}$; if this is
also the absorption optical depth at 1.1 \micron, then the albedo of the
scatterers is $\sim$0.3.  However, about \onethird\ of the luminosity
emitted from 3.5 -- 25 \micron\ arises from a region within 150 AU of
the star (\cite{Silverstone99}). So, the total infrared $\tau$
overestimates the optical depth in the exterior part of the disk seen in
reflected light and results in an underestimate of the albedo.  In the
region at $\sim$1.$''$9 (190 AU), therefore, the albedo is greater than
0.3 and probably more like 0.4.  This is a substantially higher albedo
than the zodiacal dust in our own Solar system, 0.1 (\cite{Dumont88}),
but is similar to the albedos of mineral mixtures and ices found in ISO
spectra of Herbig Ae/Be disks (\cite{Malfait98}).

At r$<$150 AU, the dust temperature is $\sim$150 K as estimated from
mid-infrared colors and luminosity (\cite{Silverstone99}).  If the grains
are icy, the decline in 1.1 \micron\ surface brightness inside 185 AU
may be due to the sublimation of the ice at temperatures $\sim$100 K. In
this case, the grains emitting in the mid to far-infrared are the
iceless counterparts of the grains at higher radius.

Gaseous CO was detected by Zuckerman, Forveille, \& Kastner (1995) with a velocity width of
7.6 km~s$^{-1}$, corresponding to a radius of 86 AU given a disk
inclination of 51\degr\ and stellar mass of 2.3M$_\odot$.  The gas,
therefore, is well inside the brightest part of the reflected light disk
and the gap.  In the spectrum of Sylvester et al. (1996), a large rise in
the continuum shortward of 8 \micron\ and a possible peak at 11.3
\micron\ both indicate emission from polycyclic aromatic hydrocarbons.
The existence of gas and small grains in the presence of forces which
remove them very quickly indicates they are constantly being resupplied.
The disk seems to be in an active planetesimal building phase.

\section{Conclusions}

NICMOS images reveal a large circumstellar disk around the Herbig Ae star HD
141569.  The structure of the disk, including a region of depleted material at
250 AU can most easily be understood as due to dynamical sculpting by one or
more planets orbiting within it.  If planets exist in this system, they
must have formed in $<$10$^7$ yr and far from the central star compared with
the planets in our own Solar system.

\acknowledgements

We thank David Koerner, Andrea Ghez and Mike Jura for helpful
conversations.  This work is based on observations with the NASA/ESA
Hubble Space Telescope, obtained at the Space Telescope Science
Institute, which is operated by the Association of Universities for
Research in Astronomy, Inc. under NASA contract NAS-26555 and supported
by NASA grant NAG5-3042 to the NICMOS instrument definition team.

\figcaption[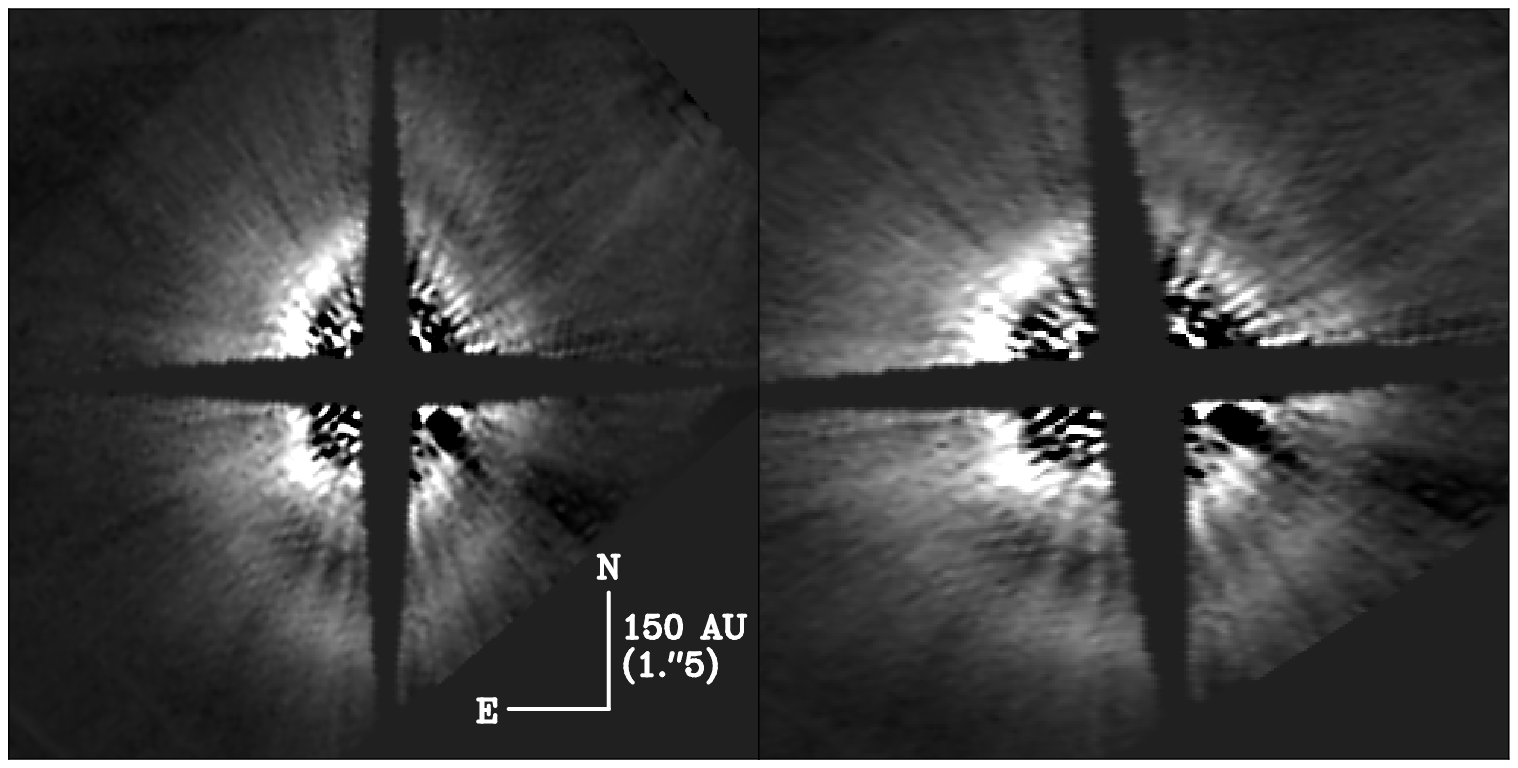]{{\it Left (a):} Image of the reflected
light disk around HD 141569 shown in a stretch which is linear with flux
density. The eastern side of the disk is brighter because of the
intrinsic scattering phase function of the dust. {\it Right (b):}
Deprojected image of the disk using an inclination angle of 51\degr.
Artifacts from small instabilities in the PSF subtraction process
include the radial ``streamers'' and the dark ``finger'' in the inner
disk at the 4 o'clock position.  These residuals do not significantly
affect the radial profile or the integrated flux density.  The
diffraction spikes narrow to higher radius as the overlap between the
two sets of observations at different orientations increases and allows
pixels behind the spikes in one image to be replaced by pixels from the
other image. \label{figimg}}
\begin{figure}[h]
\epsfig{file=Weinberger.fig1.ps,height=3in,clip=}
\end{figure}

\begin{minipage}[b]{0.46\linewidth}
\figcaption[Weinberger.fig2.ps]{Azimuthally averaged surface brightness
profile of the disk after subtraction of three different PSF stars.  The
error bars represent only the statistical uncertainty in the mean
surface brightness at each radius.  At radii greater than $\sim$160 AU
the three profiles are consistent to within these uncertainties.  The
subtractions with $\tau^1$ Eri and 49 Cet suggest that the disk flux
falls off substantially inside of 160 AU. \label{fig_radialavg}}
\bigskip

\epsfig{file=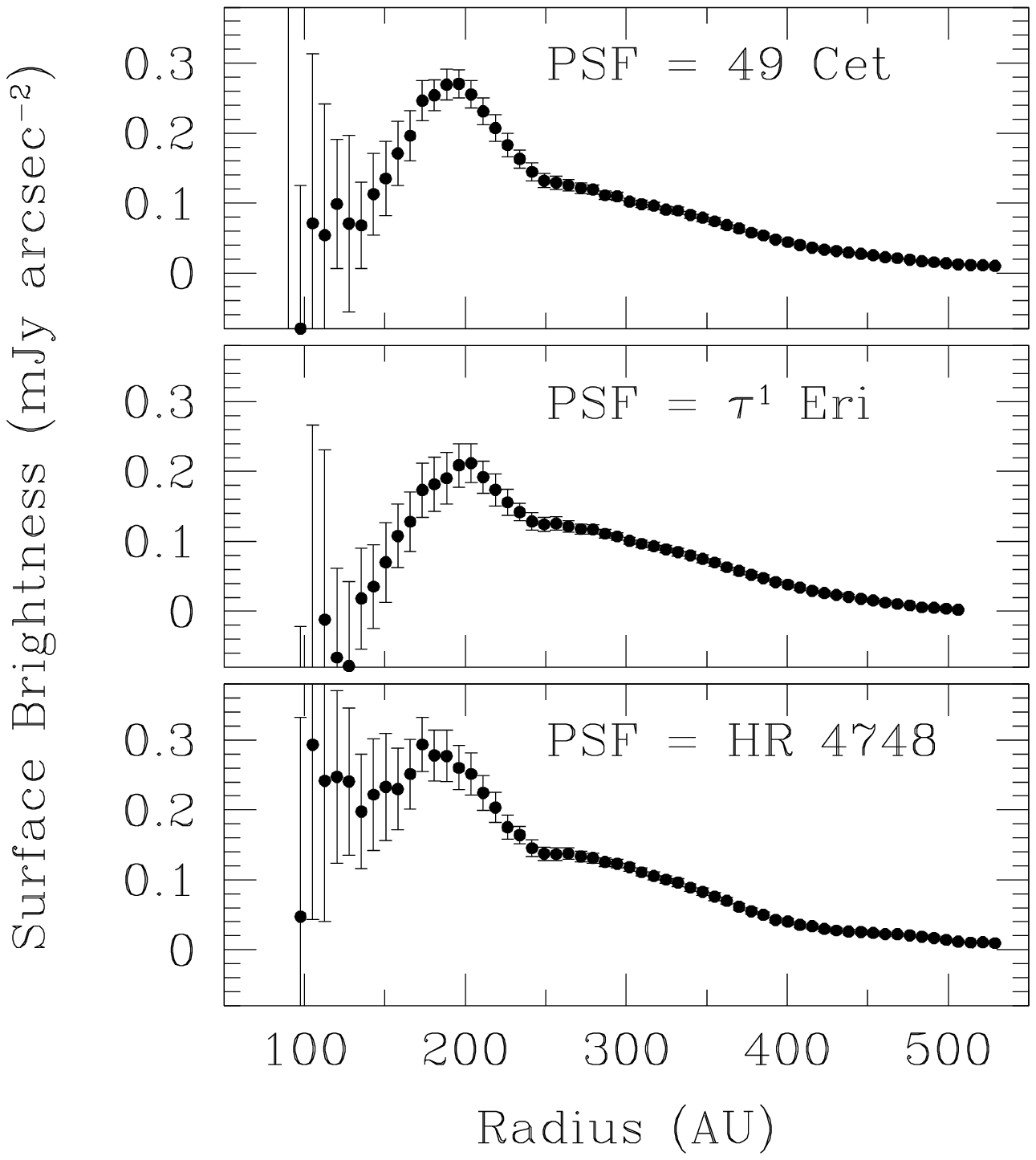,width=\linewidth,clip=}
\end{minipage}\hfill
\begin{minipage}[b]{0.46\linewidth}
\figcaption[Weinberger.fig3.ps]{The surface density of scatterers
plotted as a function of radius computed by multiplying the surface
brightness by 4$\pi$r$^2$ and normalizing to the total stellar flux
density. \label{fig_tauomega}}
\bigskip

\epsfig{file=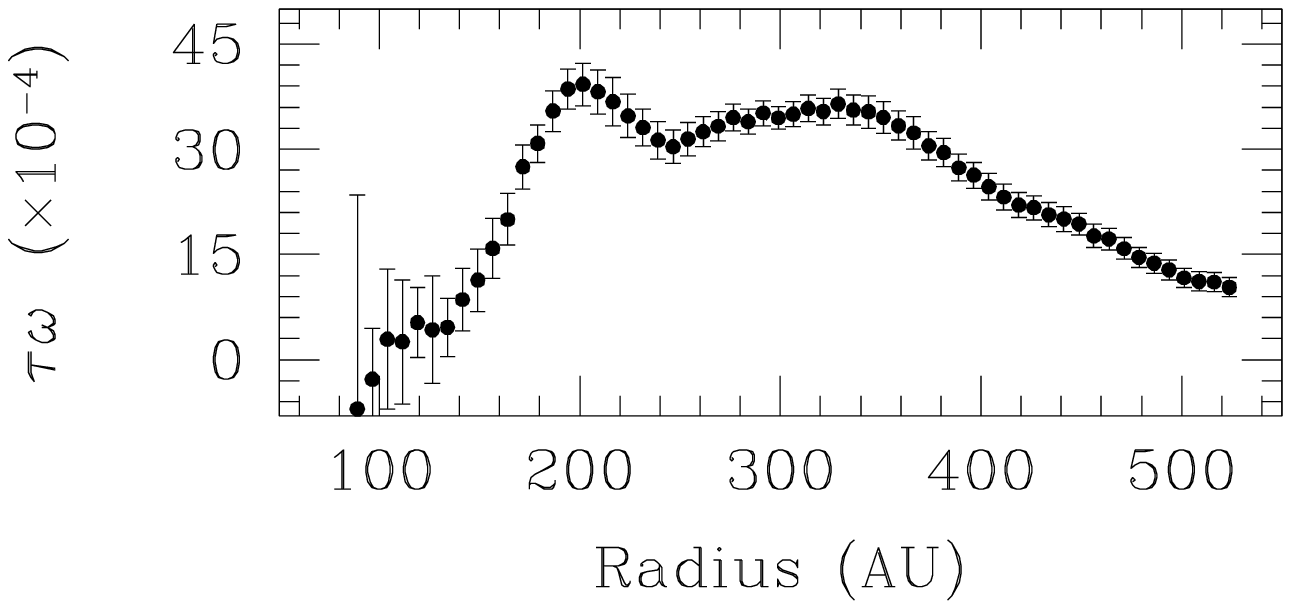,width=\linewidth,clip=}

\bigskip

\figcaption[Weinberger.fig4.ps]{Apparent F110W limiting magnitude of an
unseen companion as a function of distance from the
star. The uncertainty in the detection limit is the size of the points,
or $\sim$10\%. \label{fig_srclimit}}
\bigskip

\epsfig{file=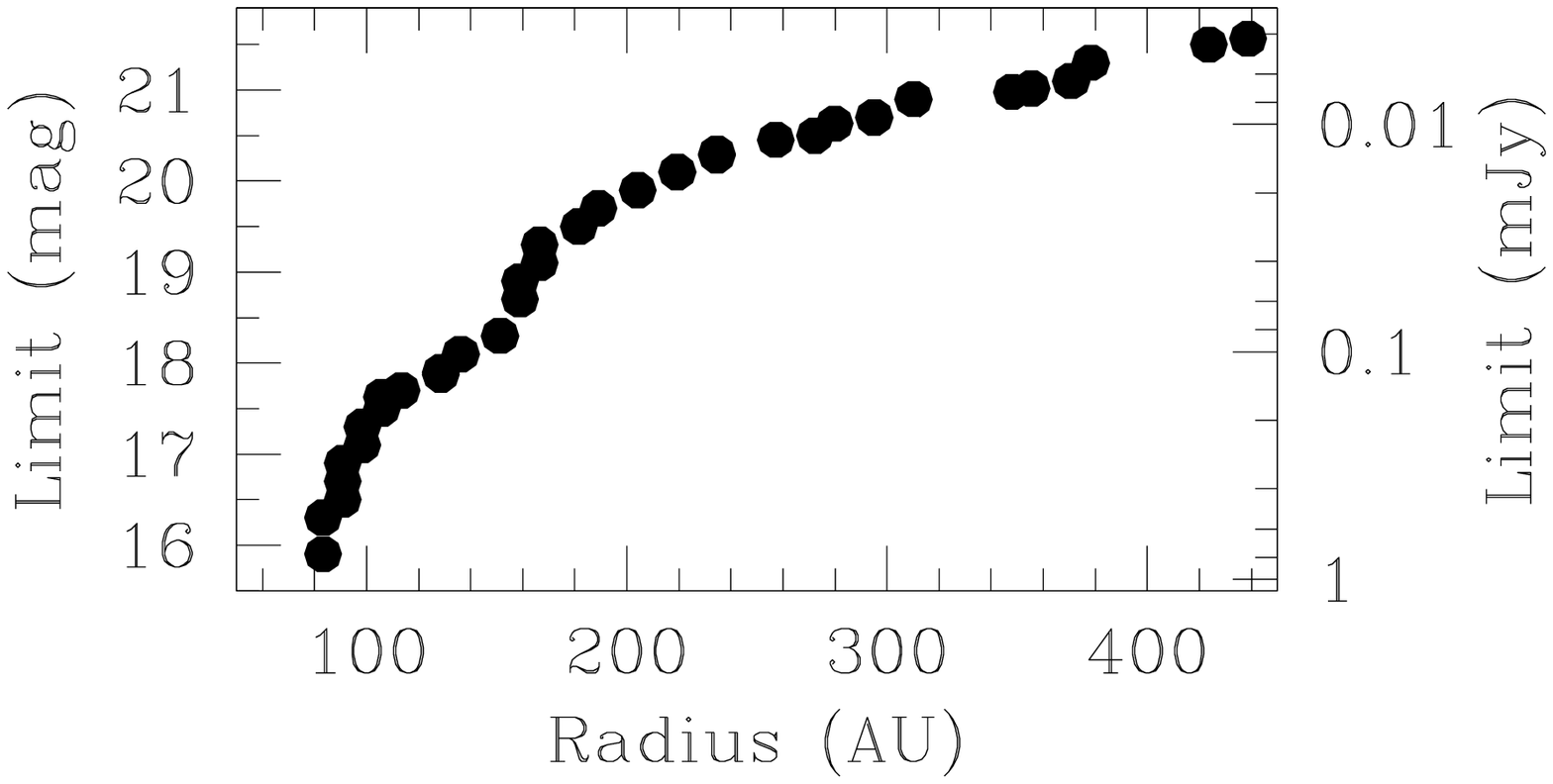,width=\linewidth,clip=}
\end{minipage}

\end{document}